\documentclass[aps,floatfix,preprint,preprintnumbers,nofootinbib,superscriptaddress,natbib]{revtex4}

\usepackage{graphicx,float}
\usepackage{epsfig}		
\usepackage{dcolumn}
\usepackage{bm}
\usepackage{subfigure}

\usepackage[all]{xy}
\usepackage{amsmath,upgreek}
\usepackage{amssymb}

\usepackage{pdfpages}
\usepackage{color}
\usepackage{graphicx,epstopdf}
\usepackage[colorlinks=true,
 linkcolor=red,
 urlcolor=purple,
 citecolor=blue,hyperindex]{hyperref}

\begin{document}

\title{Natural Capital as a Stock Option}

\renewcommand{\baselinestretch}{0}

\author{O. Bertolami}
\email{orfeu.bertolami@fc.up.pt}
\affiliation{Departamento de F\'isica e Astronomia, Faculdade de Ci\^{e}ncias da
Universidade do Porto, Rua do Campo Alegre 687, 4169-007, Porto, Portugal.}
\affiliation{Centro de F\'isica das Universidades do Minho e do Porto, Rua do Campo Alegre s/n, 4169-007, Porto, Portugal.}

\vspace{2 cm}

{\renewcommand{\baselinestretch}{1.2}

\vspace{2 cm}

\begin{abstract}
The unfolding climate crisis is a physical manifestation of the damage that market economy, driven by the high intensity consumption of fossil fuels, has inflicted on the Earth System and on the stability conditions that were established by a complex conjugation of natural factors during the Holoecene. The magnitude of the human activities and its predatory nature is such that it is no longer possible to consider the Earth System and the services it provides for the habitability of the planet, the so-called natural capital, as an economical externality. Thus one is left with two main choices in what concerns the sustainability of the planet's habitability: radical economic degrowth or highly efficient solutions to internalise the maintenance and the restoration of ecosystems and the services of the Earth System. It is proposed that an interesting strategy for the latter is to consider the natural capital as a stock option. 
 
\end{abstract}

\keywords{Earth System - Natural capital - Stock option}

\date{\today}
\maketitle

\renewcommand{\baselinestretch}{1.2}


\noindent
1. The evidence of the systemic disruption of the Earth System (ES) is overwhelming. It is manifested itself through the widespread and longer periods of droughts, heat waves, flooding, wild fires and other deleterious climate events that were once rare and are now the norm and tend to become even more frequent.These are symptoms of a serious misuse and/or overuse of the resources provided by the multiple components of the ES and the degradation of its main regulatory large scale ecosystems due to the ever increasing expansion of the human activities. The most visible result of this disruption is the change in the climate patterns due to the accumulation of anthropogenic  greenhouse gases in the atmosphere. In fact, multiple studies and the accumulated data show that we are living a climate emergency with unpredictable implications for the future generations \cite{IPCC23}. 

The international community is well aware of the seriousness of the problem and on various international climate summits the necessity to decarbonise economic activities has been acknowledged and some important agreements have been reached in order to establish targets and engage in long term commitments. Nevertheless, the implementation of these agreements has been problematic to say the least. We face a civilisational challenge and at the source of the problem are 
the basic working principles of the consumption society driven by cheap fossil fuels and built upon the mistaken assumptions that Earth's resources are infinite and that it can absorb a limitless amount of waste. These assumptions violate the most obvious principles of common sense and thermodynamics, but are the working rules of the continuous expansion of the human activity. The magnitude and the fundamental nature of the problem at hand, suggest any set of solutions for a long term fixing of the climate change crisis is not realistic as it involves no less than a dramatic reduction of the emissions of greenhouse gases and encompassing socio-economic changes.  Furthermore, it is important to realise that the problem must be addressed in little more than a decade or so and that we may be running out of time to carry out the absolutely necessary long term transformations. Given the current situation, adaptation and mitigation strategies are in the category of the absolutely minimal necessary measures in order to get us some extra time to fix the problem. These measures include accelerating efforts to decarbonise human activities, extending and generalising the use of renewable energies, setting up means to carbon capture by afforestation, restoration of ecosystems and other chemical-mechanical means, besides rational use of water and of vital natural resources. Given the urgency of the situation, even radical solutions involving geoengineering measures have been discussed, ranging from deflection of solar radiation by aerosols to removal of $CO_2$ from the atmosphere using the space lift (see, for instance, Refs. \cite{OB23,OB24} and references therein). 

Of course, as there is consensus that a drastic decarbonisation of the human activities is required, a sensible way to address the problem involves a significant reduction of the consumption patterns of our society as well as a coupled effort to change the brutal and dysfunctional way the expansion of the market economy destroys ecosystems. Thus a realistic path towards a sustainable future asks for a conscious and well planned economic degrowth. Of course, such a drastic change in the underlying ideology of the contemporary civilisation is beyond reach under the prevailing socio-economic and political conditions of our world nowadays. 

Another alternative involves renewed and vigorous forms of internalising the regulatory ecosystems into the main stream workings of the economic activities. The discussion on how this transformation might take place has been a matter of heated debate, but no effective workable solution has emerged. The idea of a tax to cover a negative externality or a market failure, a Pigovian tax, goes back to 1970s in what concerns $CO_2$ emissions. Early versions of an integrated model of climate and economy appeared in late 1970s and in more detail in 1990s through the work of the Nobel laureate economist William Nordhaus. More recent versions of this model, DICE (Dynamic Integrated Climate-Economy mode) and RICE (Regional Integrated Climate-Economy model), are a matter of intensive research and adaptation \cite{Nordhaus1,Nordhaus2}. Back in 2006, the so-called Stern report  proposed a different approach and called attention to the need to tackle the climate crisis, which was then still incipient, through a sizeable inversion of capital, about 1 to 2$\%$ of the developed countries GNP \cite{Stern1,Stern2}. This warning was dismissed and we are left today with the existing carbon social tax, which has become just another instrument of the market economy with little impact on the consumption of fossil fuels and in reducing the existing $CO_2$ in the atmosphere.  

The present state of affairs is such that after the failure of the major international agreements and commitments, Kyoto, Paris, annual COPs, adaptation and mitigation strategies are the only realistic alternatives to motivate a deeply divided international community. Under the present conditions is unlikely that any coherent and determined policy will emerge in order to face the unfolding crisis. Resilience has emerged as a basic concept and its multiple dimensions have been extensively discussed (see, for instance, Refs. \cite{Cutter1,Cutter2}). The need to incorporate the resilience factor into DICE-type models have been discussed in Refs. \cite{OBG23a,OBG23b} and a resilience social tax upon consumption has been proposed in Refs. \cite{OB22,OBG23a}.  

In the present work we argue that given the difficulties discussed above, one should consider direct forms of restoring ecosystems that have themselves an economic and a financial dimension. This involves relating the improvement of the ES performance so to ensure weather stability and sustainability as a capital asset whose value evolves according the rules of the market. This amounts to create financial assets directly connected to the maintenance and improvement of the services provided by the ES aligning a positive performance of these financial assets with actions that promote a sustainable habitability of the planet.  

Notice that this goes beyond the tenets of the usual green or circular economy as it directly relates the improvement of the ES services by assigning to them a market value. This follows the logic of the proposal of evaluating the world ecosystem services as a capital, the natural capital \cite{Costanza}.  Given the direct impact of ecosystem's restoring activities on Earth's life-supporting system and human welfare, it can have deeper implications than  improvements on the existing production processes, recycling and waste management, which are of course greener 
than the tradicional methods, but are far away from being an encompassing and effective solution for decarbonising the economy.

\vspace{0.5 cm}

\noindent
2. In order to establish the restoration of ecosystems as an economic and financial activity, one must endow it with a suitable market characterisation. As the restoration of ecosystems is a knowledge and a capital intensive activity, it is fairly straightforward to regard it as a derivative, that is an economic and engineering activity that can be furnished in a financial instrument whose worth is supported by the value of an underlying asset. In the stock market, the way to trade a given asset so to ensure stability and equity for owners and shareholders in a well defined time framework is through the concept of a stock option.  

A stock option is a derivative that endows a trader with the right, although not the obligation, to buy or sell shares of a stock at an agreed-upon price and date. This means that an option settles an agreement contract between parties so  to have the option to sell or to buy the stock in the future at a specified price. This specified price is referred to as the strike price or the exercise price.

As an asset to be traded in the future, the value of an option, the so-called pay-off, $V$, is a function of stock price, $S$, and time, $t$, that is, $V=V(S,t)$, which is an inherently stochastic variable evaluated through suitable It\o calculus manipulations and whose behaviour in modern stock trading is described by the Black-Scholes-Merton partial differential equation \cite{BS,Merton} :
\begin{equation}
 {\partial V \over \partial t} + {1 \over 2} \sigma^2 S^2  {\partial^2 V \over \partial S^2} + r S  {\partial V \over \partial S} -rV = 0, 
 \label{BSM}
\end{equation}
where $\sigma$ is the volatility of the stock, the square root of the log of the stock prices, and $r$ the interest rate.

Suitable boundary conditions to solve the Black-Scholes-Merton equation can be settled through conditions on the buying price of an (European) option, $C(S,t)$, as follows:  $C(0,t) = 0$, for any $t$ and large $T$; $C(S,t) \to S(t)$, for large S; $C(S,T)=max(S-K,0)$, where $K(T)$ is the exercise price for $t=T$, the option time. Thus, through the change of variables, 
$\tau=T-t$, $u(x,\tau)= C(S,t)e^{r\tau}$, and $x=ln(S/K) + (r - {\sigma^2 \over 2})\tau$, the  Black-Scholes-Merton equation becomes the Fourier diffusion equation:
\begin{equation}
 {\partial u(x,\tau) \over \partial \tau} =  {\sigma^2  \over 2}  {\partial^2  u(x,\tau)  \over \partial x^2} , 
\label{Fourier}
\end{equation}
which, with the initial condition $u(x, 0) \equiv u_0(x)= K max(e^x-1, 0)$, admits the well known solution
\begin{equation}
u(x,\tau) =  {1 \over \sigma \sqrt{2 \pi \tau}} \int^{\infty}_{-\infty} u_0(y) e^{-(x-y)^2/(2\sigma^2 \tau)} dy~. 
\label{solu}
\end{equation}
It is now simple to show that a useful solution of Eq. (\ref{BSM}), 
with the initial condition $u(x, 0) \equiv u_0(x)= K max(e^x-1, 0)$, is the well-known Black-Scholes formula:
\begin{equation}
C(S,t) =  S(t) N(d_1) - K(T) e^{-r (T-t)} N(d_2)~, 
\label{solu}
\end{equation}
where
\begin{equation}
N(x) =  {1 \over \sqrt{2 \pi}} \int^{x}_{-\infty}  e^{-y^2/2} dy~,
\label{normal}
\end{equation}
is the normal distribution function,  
\begin{equation}
d_1 =  {ln(S/K) +  (r + \frac{\sigma^2}{2})(T-t) \over \sigma \sqrt{T-t}}~,
\label{d1}
\end{equation}
and 
\begin{equation}
d_2 =  d_1 - \sigma \sqrt{T-t}~. 
\label{d1}
\end{equation}
On its hand, the selling price of an option, $P(S,t)$, is obtained from the buying price of the option, $C(S,t)$, written above:
\begin{equation}
P(S,t) =  K(T) e^{-r (T-t)} N(-d_2) - S(t) N(-d_1)~. 
\label{P}
\end{equation}

One can envisage several ways to internalise the operating conditions of the ES within the wrapping of a stock option. In what follows, we shall present a suggestion that is based on the idea that the damage on ES due to human activities can be evaluated by the so-called planetary boundaries (PB) \cite{Steffen:2015}, a set of at least nine parameters (rate of biosphere loss, land system change, global fresh water use, biogeochemical flows (global Nitrogen and Phosphorus cycles), ocean acidification, atmospheric aerosol loading, stratospheric ozone depletion, climate change, chemical pollution) whose values relative to their readings at the Holocene, the period during which the ES was in a stable condition. This latter state is referred to as the safe operating space (SOS) \cite{Rockstrom:2009}. In fact, it has been argued that a PB quota-based system is the most direct way to measure the impact of the human activities on the ES \cite{Meyer}, but of course, different parameterisations can be used and the proposal that is discussed below is just an example of the approach that we suggest here. 


\vspace{0.5 cm}

\noindent
3. We have reached a key point of our discussion. The stock is a financial instrument meant to generate profit and that can be completely abstract with no contact with the material world. Our suggestion is to attach a stock to the improvement of the ES via the PB parametrisation. This parametrisation is an evaluation element of a particular action on the environment and on the ES with a direct impact on the natural capital. Thus, our suggestion can be summarised in a simple principle: a stock can be created so that its price yields a profit if it adds up to the natural capital and allows for the improvement of the services the ES provides to the habitability of the planet. 

To proceed in the discussion we need to specify the features of the PB parameters and how they can be used to set up the price of a stock. The Great Acceleration of the human activities in 1950s \cite{Steffen:2014} has led the ES to be predominantly driven by human activities. This dominance defines the Anthropocene. If we denote the human drivers, collectively by $H$, it is possible to consider, through a suitable model \cite{Bertolami:2018,Bertolami:2019}, the impact of the human activities on the ES through the value of the PB parameters, $h_i$, relative to their values at the SOS. Thus, the most general form of $H$, allowing for interaction between the various PB parameters is given \cite{Bertolami:2019}:
\begin{equation}
H = \sum_{i=1}^{9} h_i + \sum_{i,j=1}^{9} g_{ij} h_i h_j + \ldots,
\label{eqn:human_action}
\end{equation}
where the second and the following set of terms parametrise the interactions among the various effects of the human action on the ES. Of course, higher order interactions terms can be considered, but it us usual to restrict the  discussion up to second order. It is physically sensible and mathematically convenient to assume that the $9 \times 9$ matrix, $[g_{ij}]$ is symmetric, $g_{ij} = g_{ji}$, and non-degenerate, $\det[g_{ij}] \neq 0$. It is natural to assume that these interactions terms are sub-dominating, and their importance has to be established through a suitable empirical analysis. As shown in Ref.\,\cite{Bertolami:2019}, they can lead to new equilibria and suggest some mitigation strategies depending on the sign and strength of the matrix entries, $g_{ij}$ \cite{Bertolami:2019}. In fact, this modelling was used to show that the interaction term between the climate change variable ($CO_2$ concentration), $h_1$, and the oceans acidity, $h_2$, is non-vanishing and about $10\%$ of the value of the single contributions themselves \cite{Barbosa:2020}. 

This parametrisation is crucial for assessing the state of the ES and, in particular, for the physical model proposed in Refs. \cite{Bertolami:2018,Bertolami:2019}, which assumes that the transition of the ES, for instance, from the Holocene to other stable states was similar to a thermodynamic phase transition and could be described by the Landau-Ginzburg Theory (LGT). According LGT,  the relevant thermodynamic variable required to specify the state of the ES is the free energy, $F$, and phase transitions are expressed through a relevant order parameter, $\psi$, in the case of the ES, the reduced temperature relative to the Holocene average temperature, $\langle T_{\rm H}\rangle$, $\psi = (T - \langle T_{\rm H}\rangle)/ \langle T_{\rm H}\rangle$. This framework allows for determining the state and equilibrium conditions of the ES in terms of physical variables, $(\eta, H)$, where $\eta$ are associated with the astronomical, geophysical and internal dynamical drivers, while $H$ corresponds to the human activities, introduced in the phase transition model as an external field. The so-called Anthropocene equation corresponds to the evolution equation of the ES once it is dominated by the human activities. The transition from the Holocene to the Anthropocene took place through the so-called great acceleration of the human activities, visible from the second half of the 20th century onwards \cite{Steffen:2014}. The proposed approach allows one to perform a phase space analysis of the temperature field, $\psi$, once an assumption for the evolution oh $H$ is considered. If a linear evolution hypothesis for $H$ is assumed then it is shown that the recently discussed Hothouse Earth scenario \cite{Steffen:2018} corresponds to an attractor of the trajectories of the ES dynamical system for regular trajectories\footnote{The capability of the proposed model to describe these planetary transitions suggests that it can provide also an interesting framework for the classification of rocky exoplanets and their terraforming \cite{OBFFrancisco2022}.}. 
If instead, a logist map is chosen for the behaviour of $H$, then for a sufficiently large rate of growth of the human activities, the ES can show stability point bifurcations and even chaotic behaviour in the phase space \cite{Bernardini:2022}. 

Notice that the parametrisation of the state of ES in terms of the PB is fairly general and Eq. (\ref{eqn:human_action}) stands independently of the physical model described above. It can therefore be used to set the price of a stock in terms of its purpose and performance. We propose that the simplest way to set the price of a stock whose purpose is to maintain and restore the services to common welfare of a ecosystem is to relate it with the improvement of the services of the ES in terms of the PB, that is:
\begin{equation}
S(t) = S_0 - \alpha \Delta H,
\label{S-price}
\end{equation}
where $S_0$ is an arbitrary initial stock price, $\alpha$ an arbitrary positive constant, $\Delta H= H(T)-H(t)$, and the minus sign is introduced so to make clear that the purpose of the proposed stock instrument is to lower the impact of the human activities on the ES.  Of course, one could break $H$ in its PB components and design a stock instrument based on a single or a set of PB parameters. Obviously, considering the interaction terms is a matter of choice, but given their non-negligible values and their physical significance, it is fairly natural to consider them. Moreover, it is clear that the strike price, $K(T)$, is set upon the performance with respect to the PB parameters. If so, it would have a price for $t>T$ set by an equation like Eq. (\ref{S-price}).  

It is then evident that the improvement of natural capital consists in diminishing the impact of the human activities, that is decreasing H or its components, if interactions are such that the entries of $g_{ij}$ are positive. This will increase the price, $S(t)$, and the buying prize, $C(S,t)$, of the asset, and hence making its selling more attractive, which, on its turn, further boosts actions to improve the workings of the ES. Of course, more involved financial strategies could be considered and devised. To further illustrate this point let us consider two simple examples:

\noindent
i) Direct capture of $CO_2$ in the atmosphere, which lowers the climate change variable, say $h_1(t)$, so that $\Delta H < 0$ and through Eq. (\ref{S-price}) this action boosts the price of the option to higher values; 

\noindent
ii) The somewhat more involved case of the interaction between climate change variable, $h_1(t)$ and the oceans acidity, $h_2(t)$ studied in Ref. \cite{Barbosa:2020}. Oceans absorb atmospheric $CO_2$ which is turned in carbonic acid, thus raising their acidity. To maximise the price $C(S, t)$ the following conditions must be satisfied at the extremum $(h_1^0, h_2^0)$ \footnote{The general case requires studying the eigenvalues and the determinant of the Hessian matrix, $(H_S)_{i,j} = \left({\partial^2 S \over \partial h_i \partial h_j}\right)$.}:
\begin{equation}
{\partial^2 S (h_1^0, h_2^0) \over \partial h_1^2} {\partial^2 S (h_1^0, h_2^0) \over \partial h_2^2}  - \left({\partial^2 S (h_1^0, h_2^0) \over \partial h_1 \partial h_2}\right)^2 > 0 ,
\label{Hessian1}
\end{equation}
and 
\begin{equation}
{\partial^2 S (h_1^0, h_2^0) \over \partial h_1^2} < 0~,
\label{Hessian2}
\end{equation}
meaning that restoring the capacity of the oceans to absorb atmospheric $CO_2$ by lowering their acidity is a profitable action. Needless to say that a task of this sort would have on a global scale a huge positive impact on marine ecosystems and on the overall state of the ES.

We stress once again that our proposal concerning the connection between the maintenance and improvement of the natural capital and the services it provides and the stock price as an option can be achieved through any other parametrisation system to evaluate the former. In fact, for this purpose, one could very well consider any ecological footprint signature of the human activity on a given component of the ES.  

\vspace{0.5 cm}

\noindent
4. In this work we have argued that an interesting strategy to internalise the natural capital is through the creation of financial instruments with direct impact on the ES through the restoration of ecosystems. enterprises to restore ecosystems already exist, but to our knowledge their activities are not profit oriented and are not funded, as we suggest, by financial instruments that materialise in the stock market as options. This endows the somewhat abstract concept of natural capital as a palpable reality in the financial world allowing for a direct internalisation of the services of the ES to the habitability conditions of the planet. Given that the global and international nature of the market is an unchallenged reality and that its expansion is much more effective than its regulation through international climate and environment agreements, the internalisation of the ES as we propose might hopefully have a relevant impact. Thus, if the restoration of the ES can be successfully internalised via a profitable financial asset according to the very rules of the market, its impact can be complementary to the one of the international climate agreements. If an economic and financial instrument can be evaluated through parameters of the ES as suggested by Eq. (\ref{S-price}), then  its financial success will be aligned with the maintenance and improvements of the life-supporting services of the ES, in opposition to the usual market activities which are profitable through the destruction or the dumping of waste on the existing ecosystems. We believe that this can be a relevant change in paradigm. Any attempt to make the restoration of ecosystems a profitable business is helpful to fix the climate change crisis, which according to some is the biggest market blunder ever. Another potential benefit of the proposal discussed here is that it may be an additional incentive to ecosystem restoring business at community level, a fundamental cornerstone of the mitigation and adaptation activities \cite{OBG21}, meaning that these activities can be advantageously coupled together. Furthermore, putting  forward a financial asset based on a real world action of restoring ecosystems requires the conjugation of a wide range of skills from scientific to administrative, from engineering to financial and planing ones, which, on its own, helps in creating the ground for the multi-dimensional approach that is required to design solutions for coping with the climate change crisis. 

For sure, none of the exposed hinders the need for stewardship measures to rationale manage the ES. The discussion about the need for stewardship measures goes back to the very discussion about the onset of the Anthropocene itself as a new geological epoch whose habitability conditions of the planet are conditioned by the extent of the human activities \cite{Steffen:2011,Steffen:2018}. In fact, it is widely believed that some form of stewardship is necessary given the fragmented nature of the international juridical order and the urgency of the climate crisis. In the opposite direction, it has been argued that stewardship measures would be an obstacle to the expansion of the free market hindering its benefits and its underlying liberal values. Given that objective historic developments do not support these claims and that international economic and financial crises arise quite regularly, it is no longer possible to hide the need for a new international economic order to prevent an irreversible planetary climate catastrophe. Furthermore, since market economy cannot avoid widespread poverty and deep social inequalities, it cannot, on its own, avoid the dysfunctionalities  it creates. These dysfunctionalities link together social inequalities and climate crisis \cite{OB22}, Thus, stewardship measures may function as regulatory boundaries which can be exerted through the form of a new ES justice \cite{Gupta} or via decentralised means based, for instance, on digital contracts \cite{OB:2020}. In any case, it is clear that concerted and focussed action is required to face the ongoing climate crisis and the social challenges it ensues. Internalising the life-supporting services provided by the ES does not address the main flaws of the market economy, but it might render an alternative to the more logical, although unfeasible economic degrowth. Therefore, devising ways to introduce the natural capital into the financial market in order to fund the specialised task of restoring ecosystems appears to be a fairly sensible approach, wrapping these restoring activities as a derivative like a stock option seems just natural.



\end{document}